\documentclass[pra,showpacs,twocolumn,amsmath]{revtex4}
\usepackage{epsfig}

\def\ra{\rangle}
\def\la{\langle}

\def\Cb{{\Bbb C}}
\def\be{\begin{equation}}
\def\ee{\end{equation}}
\def\ba{\begin{array}}
\def\ea{\end{array}}
\def\Cb{\ \hbox{\vrule width 0.6pt height 6.5pt depth 0pt
              \hskip -3.2 pt} C}
\begin{document}

\title{Experimental Determination of Entanglement for Arbitrary Pure States}
\author{Shao-Ming Fei$^{1,2}$}
\author{Ming-Jing Zhao$^{1}$}
\author{Kai Chen$^{3}$}
\author{Zhi-Xi Wang$^{1}$}

\affiliation{$^1$School of Mathematical Sciences, Capital Normal
University, Beijing 100048,
China\\
$^2$Max-Planck-Institute for Mathematics in the
Sciences, 04103 Leipzig, Germany\\
$^3$Hefei National Laboratory for Physical Sciences at Microscale
and Department of Modern Physics, University of Science and
Technology of China, Hefei, Anhui 230026, China}

\begin{abstract}
{We present a way of experimentally determining the concurrence in
terms of the expectation values of local observables  for arbitrary
multipartite pure states. In stead of the joint measurements on two
copies of a state in the experiment for two-qubit systems [S. P.
Walborn \emph{et al}. Nature (London)\textbf{440}, 20(2006)], we
only need one copy of the state in every measurement for any
arbitrary dimensional multipartite systems, avoiding the preparation
of twin states or the imperfect copy of the state.}
\end{abstract}

 \maketitle

\section{INTRODUCTION}
Quantum entanglement is one of the most fascinating features of
quantum theory \cite{M.A.Nielsen}. To characterize and quantify the
entanglement some well defined measures such as entanglement of
formation (EOF) \cite{eof,Horo-Bruss-Plenioreviews} and concurrence
\cite{concurrence,anote} have been used. For given two-quibt or some
special symmetric states explicit analytic formulas for EOF and
concurrence have been found
\cite{wotters,Terhal-Voll2000,fjlw,fl,fwz,Rungta03}. For arbitrary
given states the entanglement can be estimated by analytic lower
bounds
\cite{167902,Chen-Albeverio-Fei1,chen,breuer,breuerprl,vicente,zhang,gao,edward,ou}.

Nevertheless, for unknown quantum states, to characterize the
entanglement one needs experimental measurements. An important
approach to detect entanglement is the Bell-type inequalities
\cite{Bell,CHSH,Gisin,Jing-Ling,sixiayu}. For instance, Gisin proved
that all two-qubit pure entangled states violate the
Clauser-Horne-Shimony-Holt (CHSH) inequality \cite{Gisin} and Chen
{\sl et al.} presented a Bell-type inequality that would be violated
by all three-qubit pure entangled states \cite{Jing-Ling}. For
general mixed two-qubit states, Yu {\sl et al.} \cite{sixiayu}
proposed a Bell-type inequality that gives a sufficient and
necessary criterion for separability. Another experimentally
plausible approach is the entanglement witness \cite{winess}, which
could also be used to detect certain kinds of entangled states with
the present technology \cite{M. Bourennane}.

However, to detect the entanglement in terms of Bell-type
inequalities one needs expectation values of two or more observables
(two or more setting measurements) per party. And one has to do
infinitely many dichotomic measurements theoretically. Moreover
untill now we still have no necessary and sufficient Bell
inequalities to detect the entanglement for general multiqubit
systems. Certain entanglement witness works only for some special
states.

In fact the concurrence is defined for both bipartite and
multipartite states and gives rise to not only the separability, but
also the degree of entanglement (at least for arbitrary dimensional
bipartite states). The problem is how to use this measure to
determine the entanglement for unknown quantum states
experimentally.

In \cite{Florian Mintert} Mintert {\sl et al.} proposed a method to
measure the concurrence directly by using joint measurements on two
copies of a pure state. Then Walborn {\sl et al.} presented an
experimental determination of concurrence for two-qubit states
\cite{S. P. Walborn,S. P. Walborn2007}, where only one-setting
measurement is needed, but two copies of the state have to be
prepared in every measurement.

In this paper, we give a way of experimental determination of
concurrence for two-qubit and multi-qubit states, such that only
one-copy of the state is needed in every measurement. To determine
the concurrence of the two-qubit state used in \cite{S. P.
Walborn,S. P. Walborn2007}, also one-setting measurement is needed,
which avoids the preparation of the twin states or the imperfect
copy of the unknown state, and the experimental difficulty is
dramatically reduced. As examples general two-qubit and three-qubit
systems, and generalized multiqubit GHZ states $|\psi\rangle=a_0 |0
\cdots 0 \rangle +|1\cdots 1\rangle$ are investigated explicitly.
The results are then generalized to the case of arbitrary
dimensional multipartite pure states.

\section{CONCURRENCE FOR N-QUBIT SYSTEM}

For a $N$-partite $M$ dimensional pure state
$|\psi\ra=\sum_{i_1,~\cdots,~i_N=0}^{M-1}a_{i_1,~\cdots,~i_N}|i_1,~\cdots,~i_N\ra$,
$a_{i_1,~\cdots,~i_N}\in \Cb$, the concurrence is given by \cite{L.
Aolita},
\begin{eqnarray}
\label{multiconred}
C(|\psi\ra)=2^{1-\frac{N}{2}}\sqrt{(2^N-2)-\sum_i tr\rho_i^2},
\end{eqnarray}
where the summation goes over all $2^N-2$ subsets of the $N$
subsystems, $\rho_i$ is the corresponding reduced density matrix
with respect to the $i$th bipartite decomposition.

Up to a constant factor, $C(|\psi\ra)$ can also be written as
\cite{anote},
\begin{eqnarray}\label{multicon1}
C(|\psi\ra)=\sqrt{\sum_p\sum_{\{\alpha,~\alpha^\prime,~\beta,~\beta^\prime\}}^M
|a_{\alpha\beta}a_{\alpha^\prime\beta^\prime}-a_{\alpha\beta^\prime}a_{\alpha^\prime\beta}|^2},
\end{eqnarray}
where $\alpha$ and $\alpha^\prime$ ($\beta$ and $\beta^\prime$) are
subsets of the subindices of $a$, associated with the same
sub-Hilbert spaces but with different summing indices. $\alpha$
($\alpha^\prime$) and $\beta$ ($\beta^\prime$) span the whole space
of a given subindex of $a$. $\sum_p$ stands for the summation over
all possible combinations of the indices of $\alpha$ and $\beta$.

Our main aim is to re-express the concurrence in terms of the expectation values of local
observables. We first give a general proof that
this can be always done: the squared concurrence of $N$-qubit pure state
$|\psi\ra$, $C^2(|\psi\ra)$, can be expressed by the real linear
summation of $\la\psi|
\sigma_{i_1}\sigma_{i_2}\cdots\sigma_{i_N}|\psi\ra \la\psi|
\sigma_{j_1}\sigma_{j_2}\cdots\sigma_{j_N}|\psi\ra$,
\begin{equation}\label{2*...*2}
\begin{array}{rcl}
C^2(|\psi\ra)&=&\displaystyle\sum_{i_1,~\cdots,~i_N,~j_1,~\cdots,~j_N=0}^3x_{i_1,~\cdots,~i_N,~j_1,~\cdots,~j_N}\\[6mm]
&&\la\psi|\sigma_{i_1}\sigma_{i_2}\cdots\sigma_{i_N}|\psi\ra \la\psi|
\sigma_{j_1}\sigma_{j_2}\cdots\sigma_{j_N}|\psi\ra,
\end{array}
\end{equation}
where the coefficients $x_{i_1,~\cdots,~i_N,~j_1,~\cdots,~j_N}$ are
real, $\sigma_{0}$ is the $2\times 2$ identity matrix, $\sigma_1=
\left(
\begin{array}{cc}
0 & 1  \\
1 & 0
\end{array}
\right)$, $\sigma_2= \left(
\begin{array}{cc}
0 & i  \\
-i & 0
\end{array}
\right)$ and $\sigma_3= \left(
\begin{array}{cc}
1 & 0  \\
0 & -1
\end{array}
\right)$ are the Pauli matrices.

We only need to show that each term in the squared $C(|\psi\ra)$ of (\ref{multicon1})
can be written in the form of the right hand side of (\ref{2*...*2}). Note that
\begin{equation}\label{eachterm}
\begin{array}{l}
|a_{\alpha\beta}a_{\alpha^\prime\beta^\prime}-a_{\alpha\beta^\prime}a_{\alpha^\prime\beta}|^2=
|a_{\alpha\beta}a_{\alpha^\prime\beta^\prime}|^2+|a_{\alpha\beta^\prime}a_{\alpha^\prime\beta}|^2\\[3mm]
~~~-a_{\alpha\beta}^*a_{\alpha^\prime\beta^\prime}^*a_{\alpha\beta^\prime}a_{\alpha^\prime\beta}-
a_{\alpha\beta}a_{\alpha^\prime\beta^\prime}a_{\alpha\beta^\prime}^*a_{\alpha^\prime\beta}^*.
\end{array}
\end{equation}
Set  $A^{(11)}=|\alpha\beta\ra\la\alpha\beta|$,
$A^{(21)}=|\alpha^\prime\beta^\prime\ra\la\alpha^\prime\beta^\prime|$,
$A^{(12)}=|\alpha\beta^\prime\ra\la\alpha\beta^\prime|$,
$A^{(22)}=|\alpha^\prime\beta\ra\la\alpha^\prime\beta|$, then
\be\label{nqubit1}
\ba{ll}
|a_{\alpha\beta}a_{\alpha^\prime\beta^\prime}|^2+|a_{\alpha\beta^\prime}a_{\alpha^\prime\beta}|^2&=
\la\psi| A^{(11)}|\psi\ra\la\psi| A^{(21)}|\psi\ra\\[3mm]
&+\la\psi|A^{(12)}|\psi\ra\la\psi| A^{(22)}|\psi\ra.
\ea
\ee
$A^{(ij)}$, $i,j=1,2$, obviously has the form $A^{(ij)}=|i_1\ra\la
i_1|\otimes\cdots\otimes|i_N\ra\la i_N|$, where $i_1,\cdots,i_N$
take value 0 or 1. As $|0\ra\la0|=\frac{1}{2}(\sigma_0+\sigma_3)$ and
$|1\ra\la1|=\frac{1}{2}(\sigma_0-\sigma_3)$, we have
\begin{widetext}
$$
|a_{\alpha\beta}a_{\alpha^\prime\beta^\prime}|^2+|a_{\alpha\beta^\prime}a_{\alpha^\prime\beta}|^2=
\displaystyle\sum_{i_1,\cdots,i_N,j_1,\cdots,j_N=0,3}x_{i_1,\cdots,i_N,j_1,\cdots,j_N}\la\psi|
\sigma_{i_1}\sigma_{i_2}\cdots\sigma_{i_N}|\psi\ra \la\psi|
\sigma_{j_1}\sigma_{j_2}\cdots\sigma_{j_N}|\psi\ra
$$
\end{widetext}
for some real coefficients $x_{i_1,\cdots,i_N,j_1,\cdots,j_N}$.

Denote further
$A^{(13)}=\frac{1}{\sqrt{2}}(|\alpha\beta\ra\la\alpha\beta^\prime|+|\alpha\beta^\prime\ra\la\alpha\beta|)$,
$A^{(23)}=\frac{1}{\sqrt{2}}(|\alpha^\prime\beta\ra\la\alpha^\prime\beta^\prime|
+|\alpha^\prime\beta^\prime\ra\la\alpha^\prime\beta|)$,
$A^{(14)}=\frac{i}{\sqrt{2}}(|\alpha\beta\ra\la\alpha\beta^\prime|-|\alpha\beta^\prime\ra\la\alpha\beta|)$,
$A^{(24)}=\frac{i}{\sqrt{2}}(|\alpha^\prime\beta\ra\la\alpha^\prime\beta^\prime|
-|\alpha^\prime\beta^\prime\ra\la\alpha^\prime\beta|)$,
then
\begin{widetext}
\begin{eqnarray}\label{nqubit2}
-a_{\alpha\beta}^*a_{\alpha^\prime\beta^\prime}^*a_{\alpha\beta^\prime}a_{\alpha^\prime\beta}-
a_{\alpha\beta}a_{\alpha^\prime\beta^\prime}a_{\alpha\beta^\prime}^*a_{\alpha^\prime\beta}^*=-
(\la\psi| A^{(13)}|\psi\ra\la\psi| A^{(23)}|\psi\ra+\la\psi|
A^{(14)}|\psi\ra\la\psi| A^{(24)}|\psi\ra).
\end{eqnarray}
\end{widetext}
It is clear that $|\alpha\ra\la\alpha|$,
$|\alpha^\prime\ra\la\alpha^\prime|$, $|\beta\ra\la\beta^\prime|$
and $|\beta^\prime\ra\la\beta|$ are tensor products of
$|0\ra\la0|$, $|1\ra\la1|$,
$|0\ra\la1|=\frac{1}{2}(\sigma_1-i\sigma_2)$ and
$|1\ra\la0|=\frac{1}{2}(\sigma_1+i\sigma_2)$. Without loss of generality
we assume $A^{(13)}=|i_1\ra\la i_1|\otimes\cdots\otimes|i_s\ra\la
i_s|\otimes(|i_{s+1}\ra\la j_{s+1}|\otimes\cdots\otimes|i_N\ra\la
j_N|+|j_{s+1}\ra\la i_{s+1}|\otimes\cdots\otimes|j_N\ra\la i_N|)$,
where $1\leq s<N$, $i_k,j_k$ take values 0 or 1 and $i_k\neq j_k$ for
each $s+1\leq k\leq N$. The part $|i_1\ra\la
i_1|\otimes\cdots\otimes|i_s\ra\la i_s|$ is the real linear
summation of tensor products of $\sigma_0$ and $\sigma_3$. While the
rest part $T\equiv|i_{s+1}\ra\la
j_{s+1}|\otimes\cdots\otimes|i_N\ra\la j_N|+|j_{s+1}\ra\la
i_{s+1}|\otimes\cdots\otimes|j_N\ra\la i_N|$ can be written as
$$
\frac{1}{2^{N-s}}\bigotimes_{l=s+1}^N(\sigma_1+i(-1)^{p_l}\sigma_2)+\frac{1}{2^{N-s}}\bigotimes_{l=s+1}^N
(\sigma_1+i(-1)^{1-p_l}\sigma_2),
$$
where $p_l$ takes values 0 or 1 for each $l$. $T$ is further of the form
\begin{widetext}
$\displaystyle\frac{1}{2^{N-s}}\sum_{l=0}^{N-s}\sum_{\{No.~ l~ of~
h_j ~is ~1, ~the~ others~ are~
2\}}i^{N-s-l}\bigotimes_{j=1}^{N-s}\sigma_{h_j}((-1)^{l_m}+
(-1)^{N-s-l-l_m})$,
\end{widetext}
$0\leq l_m \leq N-s-l$. If $N-s-l$ is even, then $i^{N-s-l}$ is real
and each factor of $\bigotimes_{j=1}^{N-s}\sigma_{h_j}$ is
real. If $N-s-l$ is odd, then $(-1)^{l_m}+ (-1)^{N-s-l-l_m}=0$.
Hence $A^{(13)}$ is the real linear
summation of the tensor products of $\sigma_i$, $0\leq i\leq3$.

Similarly one can show that $A^{(14)}$, $A^{(23)}$ and $A^{(24)}$
are real linear summation of tensor products of $\sigma_i$, $0\leq
i\leq3$. Thus
$-a_{\alpha\beta}^*a_{\alpha^\prime\beta^\prime}^*a_{\alpha\beta^\prime}a_{\alpha^\prime\beta}-
a_{\alpha\beta}a_{\alpha^\prime\beta^\prime}a_{\alpha\beta^\prime}^*a_{\alpha^\prime\beta}^*$
and Eq. (\ref{eachterm}) can be expressed in the form of real linear
summation of
$\la\psi|\sigma_{i_1}\sigma_{i_2}\cdots\sigma_{i_N}|\psi\ra \la\psi|
\sigma_{j_1}\sigma_{j_2}\cdots\sigma_{j_N}|\psi\ra$.

Therefore the squared concurrence of $N$-qubit pure states $C^2(|\psi\ra)$ can be
expressed as the expectation values of tensor products of
$\sigma_i~(0\leq i\leq3)$, though
such expressions may be not unique (From (\ref{nqubit1}) and (\ref{nqubit2})
one sees that it is possible to
find an expression that is invariant under the permutations of the $N$ observables).

A. Concurrence for two-qubit system

For any two-qubit state $|\psi\ra
=a_{00}|00\ra+a_{01}|01\ra+a_{10}|10\ra+a_{11}|11\ra$,
$|a_{00}|^2+|a_{11}|^2+|a_{10}|^2+|a_{01}|^2=1$,
\begin{eqnarray}\label{2*21}
C^2=4|a_{00}a_{11}-a_{01}a_{10}|^2,
\end{eqnarray}
which can be expressed as
\begin{widetext}
\begin{eqnarray}\label{2*222}
C^2&=&\frac{1}{2}(1+\la\sigma_3\sigma_3\ra^2-\la\sigma_3\sigma_0\ra^2
-\la\sigma_0\sigma_3\ra^2-\la\sigma_0\sigma_1\ra^2
+\la\sigma_3\sigma_1\ra^2-\la\sigma_0\sigma_2\ra^2+
\la\sigma_3\sigma_2\ra^2).
\end{eqnarray}
\end{widetext}
Therefore for experimental determination of the concurrence, one only needs to measure $\la\sigma_3\sigma_3\ra$,
$\la\sigma_3\sigma_1\ra$ and $\la\sigma_3\sigma_2\ra$ respectively. One may also find
alternative expressions with symmetry under the exchange of the two qubits
\footnote{For instance,
$C^2=\frac{1}{16}(2+2\la\sigma_3\sigma_3\ra^2-2\la\sigma_3\sigma_0\ra^2
-2\la\sigma_0\sigma_3\ra^2-\la\sigma_0\sigma_1\ra^2-
\la\sigma_1\sigma_0\ra^2+\la\sigma_3\sigma_1\ra^2+\la\sigma_1\sigma_3\ra^2
-\la\sigma_0\sigma_2\ra^2-\la\sigma_2\sigma_0\ra^2+
\la\sigma_3\sigma_2\ra^2+\la\sigma_2\sigma_3\ra^2)$.}.

For states in Schmidt decomposition, $|\psi\ra =
a_0|00\ra+a_1|11\ra$, $|a_{0}|^2+|a_{1}|^2=1$, we have
\begin{eqnarray}\label{2*2sch}
C^2=\frac{1}{8}(1+\la\sigma_3\sigma_3\ra^2-\la\sigma_0\sigma_3\ra^2-\la\sigma_3\sigma_0\ra^2).
\end{eqnarray}
In this case experimentally we only need to measure
$\la\sigma_3\sigma_3\ra$, or simply count the probability $P(++)$,
$P(--)$ of projections $|++\ra\la++|$, $|--\ra\la--|$ with
$|+\ra=\frac{1}{\sqrt{2}}(|0\ra+|1\ra)$ and
$|-\ra=\frac{1}{\sqrt{2}}(|0\ra-|1\ra)$ respectively, since
$C^2=16P(++)P(--)$. For the state $\alpha|01\ra+\beta|10\ra$ used in
\cite{S. P. Walborn}, it is also true that only one-setting
measurement is needed. But here we only need one
copy of the state in every measurement, while in \cite{S. P. Walborn} joint measurements on
two copies of the state are needed in every measurement.

For small deviation
$|\psi^\prime\ra=\sqrt{1-\epsilon}|\psi\ra+\sqrt{\epsilon}|\phi\ra$
from an ideal pure state $|\psi\ra$ due to imperfect preparation,
where $\epsilon\in {{I\!\! R}}$ and $|\phi\ra$ is an arbitrary pure
state, our protocol shows that the concurrence obtained from the
experiment is exactly the one of $|\psi^\prime\ra$. Hence if the
parameter $\epsilon$ is small enough, the difference of the
concurrence between $|\psi\ra$ and $|\psi^\prime\ra$ would be small
enough. For a two-crystal type-I down-conversion source, with
improper spatial mode matching and spectral filtering, the imperfect
preparation procedure could result in mixed states,
$\rho=(1-\epsilon)|\psi\ra\la\psi|+\epsilon(|\alpha|^2|HH\ra\la
HH|+|\beta|^2|VV\ra\la VV|)$ instead of the ideal pure state
$|\psi\ra=\alpha|HH\ra+\beta|VV\ra$, where $H$ and $V$ stand for
horizontal and vertical linear polarizations respectively. That is,
the phase coherence between $|HH\ra$ and $|VV\ra$ is reduced by
$1-\epsilon$. Therefore the actual concurrence of $\rho$ is smaller
than that of $|\psi\ra$, $C(\rho)=(1-\epsilon)|\alpha\beta|$
\cite{S. P. Walborn2007,W. K. Wootters}. If we still measure the
state according to (\ref{2*222}) or (\ref{2*2sch}), we have
$C(\rho)=|\alpha\beta|$. Thus the relative error due to mixing is
linear in $\epsilon$.

{\it Remark}~~In principle one can always use tomography to reconstruct the
unknown state. However it requires a large number of measurements.
In particular one needs $3^N$-setting measurements to reconstruct an
arbitrary $N$-qubit density matrix. To obtain all 16
expectation values of the two-qubit density matrix, nine-setting measurements
have to be used \cite{C. F. Roos}. From (\ref{2*222})
we only need three-setting measurements to quantify the entanglement
of the state, which is much simpler than tomography approach.

B. {Concurrence for three-qubit system}

For any pure three-qubit state
$|\psi\ra=\sum_{i,j,k=0}^1a_{ijk}|ijk\ra$, the squared concurrence
is of the form,
\begin{widetext}
\begin{eqnarray}\label{2*2*21}
C^2&=&4(|a_{000}a_{111}-a_{001}a_{110}|^2+|a_{000}a_{111}-a_{010}a_{101}|^2+|a_{000}a_{111}-a_{011}a_{100}|^2+
|a_{001}a_{110}-a_{010}a_{101}|^2\\\nonumber
&&+|a_{001}a_{110}-a_{011}a_{100}|^2+|a_{010}a_{101}-a_{011}a_{100}|^2)+8(|a_{000}a_{011}-a_{001}a_{010}|^2
+|a_{000}a_{101}-a_{001}a_{100}|^2\\\nonumber
&&+|a_{000}a_{110}-a_{010}a_{100}|^2+
|a_{001}a_{111}-a_{011}a_{101}|^2+|a_{010}a_{111}-a_{011}a_{110}|^2+|a_{100}a_{111}-a_{101}a_{110}|^2).
\end{eqnarray}
\end{widetext}
Up to a constant factor, $C^2$ can be expressed as
\begin{widetext}
\begin{eqnarray}\label{2*2*22}
C^2&=&\frac{1}{4}(9-5\la\sigma_0\sigma_3\sigma_0\ra^2
-5\la\sigma_0\sigma_0\sigma_3\ra^2-5\la\sigma_3\sigma_0\sigma_0\ra^2+
\la\sigma_0\sigma_3\sigma_3\ra^2+\la\sigma_3\sigma_3\sigma_0\ra^2
+\la\sigma_3\sigma_0\sigma_3\ra^2+3\la\sigma_3\sigma_3\sigma_3\ra^2\\\nonumber
&&-3\la\sigma_0\sigma_0\sigma_1\ra^2-
3\la\sigma_0\sigma_1\sigma_0\ra^2-3\la\sigma_1\sigma_0\sigma_0\ra^2-\la\sigma_0\sigma_3\sigma_1\ra^2
-\la\sigma_1\sigma_0\sigma_3\ra^2-\la\sigma_3\sigma_1\sigma_0\ra^2+3\la\sigma_0\sigma_1\sigma_3\ra^2\\\nonumber
&&+3\la\sigma_3\sigma_0\sigma_1\ra^2+3
\la\sigma_1\sigma_3\sigma_0\ra^2+\la\sigma_3\sigma_3\sigma_1\ra^2
+\la\sigma_3\sigma_1\sigma_3\ra^2+\la\sigma_1\sigma_3\sigma_3\ra^2-3\la\sigma_0\sigma_0\sigma_2\ra^2-
3\la\sigma_0\sigma_2\sigma_0\ra^2\\\nonumber
&&-3\la\sigma_2\sigma_0\sigma_0\ra^2-\la\sigma_0\sigma_3\sigma_2\ra^2
-\la\sigma_2\sigma_0\sigma_3\ra^2-\la\sigma_3\sigma_2\sigma_0\ra^2+3\la\sigma_0\sigma_2\sigma_3\ra^2\\\nonumber
&&+3\la\sigma_3\sigma_0\sigma_2\ra^2+3
\la\sigma_2\sigma_3\sigma_0\ra^2+\la\sigma_3\sigma_3\sigma_2\ra^2
+\la\sigma_3\sigma_2\sigma_3\ra^2+\la\sigma_2\sigma_3\sigma_3\ra^2).
\end{eqnarray}
\end{widetext}
which is invariant under the permutations of the three qubits.
For experimental determination of the concurrence for
arbitrary three-qubit states, seven quantities are
needed to be measured: $\la\sigma_3\sigma_3\sigma_3\ra$,
$\la\sigma_3\sigma_3\sigma_1\ra$, $\la\sigma_3\sigma_1\sigma_3\ra$,
$\la\sigma_1\sigma_3\sigma_3\ra$, $\la\sigma_3\sigma_3\sigma_2\ra$,
$\la\sigma_3\sigma_2\sigma_3\ra$, $\la\sigma_2\sigma_3\sigma_3\ra$.

In particular for the three-qubit generalized GHZ state,
$|\psi\ra=a_0|000\ra+a_1|111\ra$, $|a_0|^2+|a_1|^2=1$, and the
generalized $W$ state $|\psi\ra=a_0|001\ra+a_1|010\ra+a_2|100\ra$,
$|a_0|^2+|a_1|^2+|a_2|^2=1$, their squared concurrence are
$12|a_{0}a_{1}|^2$ and
$8(|a_{0}a_{1}|^2+|a_{0}a_{2}|^2+|a_{1}a_{2}|^2)$ respectively. The
concurrence of both generalized GHZ states and generalized $W$
states can be measured according to the following formula:
\begin{equation}\label{2*2*2wghz}
\begin{array}{l}
C^2=\frac{1}{4}(9-5\la\sigma_0\sigma_3\sigma_0\ra^2
-5\la\sigma_0\sigma_0\sigma_3\ra^2-5\la\sigma_3\sigma_0\sigma_0\ra^2\\
+\la\sigma_0\sigma_3\sigma_3\ra^2
+\la\sigma_3\sigma_3\sigma_0\ra^2+\la\sigma_3\sigma_0\sigma_3\ra^2+3\la\sigma_3\sigma_3\sigma_3\ra^2).
\end{array}
\end{equation}
(\ref{2*2*2wghz}) shows that for experimental determination of entanglement for
these states, one needs only one-setting measurement, $\la\sigma_3\sigma_3\sigma_3\ra$.

Similar results can be obtained for multiqubit systems such as
$N$-qubit generalized GHZ state
$|\psi\ra=a_0|0\cdots0\ra+a_1|1\cdots1\ra$, $|a_0|^2+|a_1|^2=1$,
or $N$-qubit generalized $W$ state
$|\psi\ra=a_0|0\cdots01\ra+a_1|0\cdots10\ra+\cdots+a_{N-1}|10\cdots0\ra$,
$|a_0|^2+|a_1|^2+\cdots+|a_{N-1}|^2=1$.
For instance for the generalized GHZ state,
the concurrence is $|a_0a_1|$ up to a constant.
Its squared concurrence can be expressed as follows:
\begin{widetext}
\begin{equation}
C^2=1+ \displaystyle\sum_{k, k^\prime ~is~ even}^{1\leq k,
k^\prime\leq N}\langle \sigma_3^{(i_1 \cdots i_k)} \rangle \langle
\sigma_3^{(j_1 \cdots j_{k^\prime})} \rangle - \sum_{l l^\prime,~
is~ odd}^{1\leq l, l^\prime\leq N} \langle \sigma_3^{(i_1 \cdots
i_l)} \rangle \langle \sigma_3^{(j_1 \cdots j_{l^\prime})} \rangle,
\end{equation}
\end{widetext}
where $\langle \sigma_3^{(i_1 \cdots i_k)} \rangle$ denotes the
expectation value of the local operators such that the $i_1$-th,
$\cdots$, $i_k$-th are $\sigma_3$ operators and the rest are
identities.

\section{CONCURRENCE FOR $N$-PARTITE $M$-DIMENSIONAL SYSTEM}

Besides qubit systems, our approach can be also used for arbitrary
$M$-dimensional cases. In stead of the Pauli operators, one can use
the $SU(M)$ generators as observables:
\begin{eqnarray*}
\lambda_0=\sum_{j=0}^{M-1}|j\ra\la j|,
\end{eqnarray*}
\begin{eqnarray*}
\lambda_s=\sum_{j=0}^{s-1}|j\ra\la j|-s|s\ra\la s|,~~~1\leq s\leq
M-1,
\end{eqnarray*}
\begin{eqnarray*}
 \lambda_s=|j\ra\la k|+|k\ra\la
j|,~~~s=M,\cdots,\frac{1}{2}(M+2)(M-1),
\end{eqnarray*}
\begin{eqnarray*}
 \lambda_s=-i(|j\ra\la
k|-|k\ra\la j|),~~~s=\frac{1}{2}(M+1)M,\cdots,M^2-1,
\end{eqnarray*}
where $0\leq j<k\leq M-1$. Note that
\begin{widetext}
\begin{eqnarray*}
&&|0\rangle \langle 0| = \frac{1}{M}\lambda_0 +
\frac{1}{M(M-1)}\lambda_{M-1} + \frac{1}{(M-1)(M-2)}\lambda_{M-2} +
\cdots + \frac{1}{3\times2}\lambda_2 +  \frac{1}{2}\lambda_1,\\
&&|1\rangle \langle 1| = \frac{1}{M}\lambda_0 +
\frac{1}{M(M-1)}\lambda_{M-1} + \frac{1}{(M-1)(M-2)}\lambda_{M-2} +
\cdots + \frac{1}{3\times2}\lambda_2 -  \frac{1}{2}\lambda_1,\\
&&\vdots\\
&&|M-2\rangle \langle M-2| = \frac{1}{M}\lambda_0 +
\frac{1}{M(M-1)}\lambda_{M-1}- \frac{1}{M-1}\lambda_{M-2},\\
&&|M-1\rangle \langle M-1| = \frac{1}{M}\lambda_0 - \frac{1}{M}\lambda_{M-1},
\end{eqnarray*}
\end{widetext}
and for $0 \leq j< k \leq M-1$,
$| j \rangle \langle k| =\frac{1}{2}(\lambda_{s} + i\lambda_{s^\prime})$,
$| k \rangle \langle j| =\frac{1}{2}(\lambda_{s} -
i\lambda_{s^\prime})$ for some $M \leq s \leq \frac{1}{2}(M+2)(M-1)$
and $\frac{1}{2}(M+1)M \leq s^\prime \leq M^2-1$. Similar to the
proof of $N$-qubit system, it is direct to show that the squared
concurrence of the $N$-partite $M$-dimensional pure state $|\psi\ra$ can
be expressed in terms of real linear summation of $\la\psi|
\lambda_{i_1}\lambda_{i_2}\cdots\lambda_{i_N}|\psi\ra \la\psi|
\lambda_{j_1}\lambda_{j_2}\cdots\lambda_{j_N}|\psi\ra$:
\begin{widetext}
\begin{eqnarray}\label{m*...*m}
C^2(|\psi\ra)=\sum_{i_1,~\cdots,~i_N,~j_1,~\cdots,~j_N=0}^{M^2-1}x_{i_1,~\cdots,~i_N,~j_1,~\cdots,~j_N}\la\psi|
\lambda_{i_1}\lambda_{i_2}\cdots\lambda_{i_N}|\psi\ra \la\psi|
\lambda_{j_1}\lambda_{j_2}\cdots\lambda_{j_N}|\psi\ra,
\end{eqnarray}
\end{widetext}
where $x_{i_1,~\cdots,~i_N,~j_1,~\cdots,~j_N}$ are real.

\section{CONCLUSIONS}

We have proposed a method for experimentally determining the
concurrence in terms of the expectation value of local observables,
which gives not only sufficient and necessary conditions for
separability of the quantum states, but also the relative degree of
entanglement. Moreover unlike the case of Bell-type inequalities
where measurements are needed with respect to infinitely many
observables, we need only mean value of a few observables. And in
stead of the joint measurement on two copies of the state needed in
the experiment \cite{Florian Mintert, S. P. Walborn, S. P.
Walborn2007} for two-qubit states, we need only the usual
measurements on one copy of the state in every measurement for any
arbitrary dimensional multipartite states, which dramatically
simplifies the experiment and reduces the error rates and the
imperfectness in the preparation of the states. Compared with
entanglement witnesses, for which some a priori knowledge about the
states under investigation is needed, we do not need any information
before measuring the state in experiment.

\end{document}